\documentclass[11pt]{article}
\usepackage[preprint]{acl}
\usepackage{times}
\usepackage{latexsym}
\usepackage[T1]{fontenc}
\usepackage[utf8]{inputenc}
\usepackage{microtype}
\usepackage{inconsolata}
\usepackage{graphicx}
\usepackage{amsmath,amssymb}
\usepackage{booktabs}

\title{DashAct: A Progressive Diagnostic Benchmark for GUI Agents in Interactive Dashboard Analysis}
\author{
Chuhan Zhang$^{1}$, Qi Xie$^{2}$, Ziyue Wang$^{2}$, Jianing Yin$^{1}$\\
Yunfan Zhou$^{1}$, Dazhen Deng$^{1}$\thanks{Corresponding author.}, Yingcai Wu$^{1}$\\
$^{1}$State Key Laboratory of CAD\&CG, Zhejiang University\\
$^{2}$College of Computer Science and Technology, Zhejiang University\\
Hangzhou, China\\
\texttt{\{chuhanzhang,3230100951,3230100860,yinjianing\}@zju.edu.cn}\\
\texttt{\{yf.zhou,dengdazhen,ycwu\}@zju.edu.cn}
}

\begin{document}
\maketitle
\begin{abstract}
Interactive dashboards require users to reveal and connect evidence
across stateful interactions. Although graphical user interface (GUI)
agents could automate this process, existing dashboard benchmarks
primarily report final answers or task success. They provide limited
insight into whether failures arise from maintaining the analytical
process, selecting actions, or grounding visual targets.
We introduce DashAct, a progressive diagnostic benchmark that evaluates
the same annotated dashboard interactions under systematically varied
contextual and spatial support. DashAct contains 348 human-verified
interaction trajectories with milestone dependencies and hierarchical
target annotations. Its three stages evaluate end-to-end execution,
verified-context action prediction, and target-conditioned grounding
under four visual conditions. Aligned action and spatial annotations
connect outcomes across these settings, providing diagnostic evidence
for interpreting online failures.
Experiments reveal persistent difficulties in action prediction and
fine-grained grounding, alongside model-dependent responses to
contextual and spatial support. A Qwen-based adaptation study further
improves target localization and online Milestone Coverage, illustrating
how DashAct can guide practical agent improvements.
\end{abstract}

\section{Introduction}

\begin{figure*}[t]
    \centering
    \includegraphics[width=\textwidth]{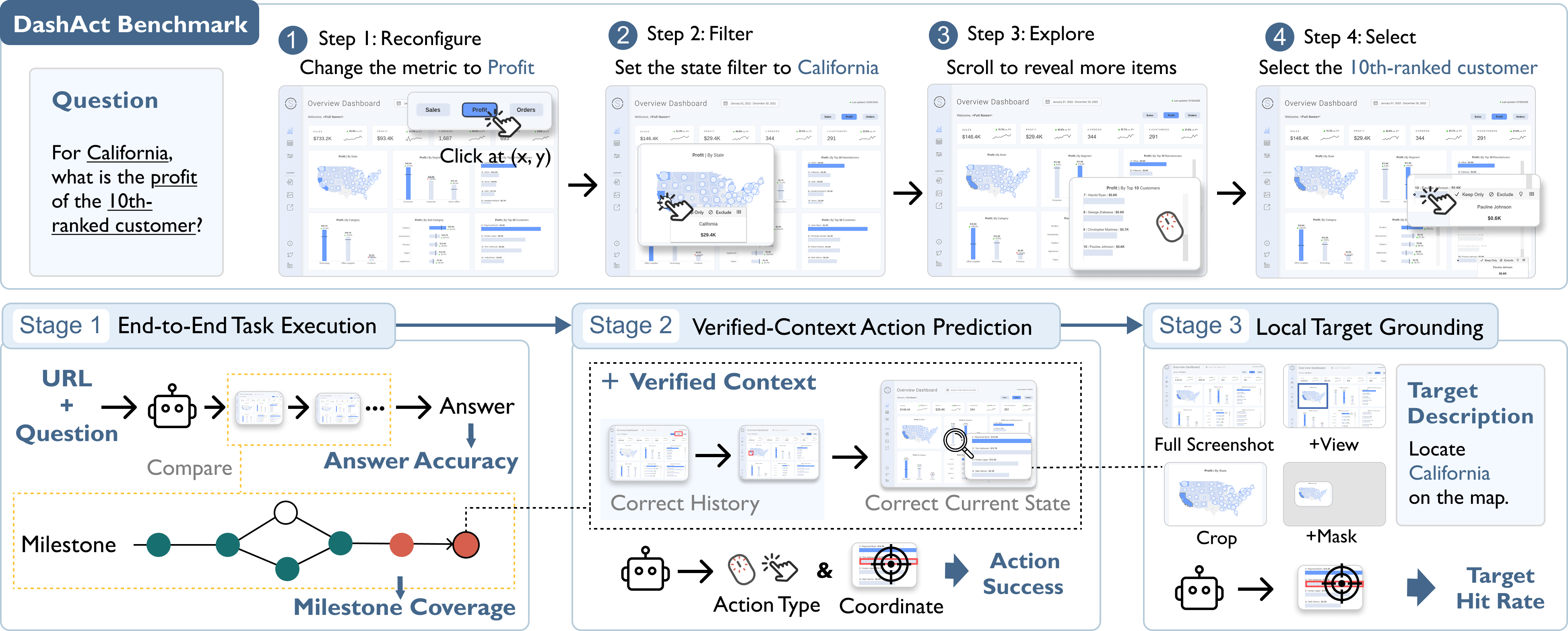}
    \vspace{-2pt}
    \caption{Overview of DashAct: the same annotated interactions are evaluated through end-to-end execution, verified-context action prediction, and target-conditioned grounding with additional spatial support.}
    \label{fig:teaser}
\end{figure*}

Multimodal foundation models have enabled graphical user interface (GUI)
agents to interpret rendered interfaces and act on user instructions~\cite{koh2024visualwebarena,xie2024osworld,cheng2024seeclick}.
While many GUI tasks focus on navigation or transactional operations,
interactive dashboard analysis requires agents to obtain and connect
analytical evidence through interactions with charts, tables, filters, and
coordinated views~\cite{yi2007toward,heer2012interactive}.
Answering a question may require configuring a view, inspecting visual marks,
and interpreting the resulting state. Interaction therefore constitutes the
analytical process through which evidence becomes available.

Dashboard interaction is challenging because actions are stateful and
interdependent. At each step, an agent must determine what information is
needed, choose an appropriate operation, locate its visual target, and
interpret the resulting interface before continuing. For example, changing
a filter or selection may update coordinated views, altering both the
evidence and actionable targets available in subsequent
steps~\cite{heer2012interactive}.
Agents must therefore track interface-state changes and retain intermediate
evidence throughout the analysis. Errors can propagate through this process
and prevent successful task completion. Meanwhile, dashboard screenshots
contain multiple views with densely arranged controls and visualization
marks, requiring agents to map intended targets to precise screen
locations~\cite{cheng2024seeclick,gou2025uground}.
These sequential and spatial demands jointly challenge autonomous dashboard
analysis.

Existing benchmarks evaluate question answering on interactive
dashboards~\cite{kartha2026dashboardqa}, track intermediate progress during
online execution~\cite{pan2024webcanvas}, and assess next-action
prediction~\cite{deng2023mind2web} and visual
grounding~\cite{cheng2024seeclick}.
However, two diagnostic challenges remain. First, milestone completion
reveals which interactions an agent performs, but a missed milestone does
not indicate whether the agent failed to reach the required interface state,
select an appropriate next action, or accurately locate its target.
Action-prediction and grounding benchmarks provide more focused assessments
of action selection and target localization, but their separate scores do
not establish how these capabilities relate to the same interactions missed
online. Second, grounding failures in multi-view dashboards may arise from
locating the relevant view or precisely identifying a fine-grained target
within it. Overall grounding accuracy does not distinguish these spatial
difficulties.

We introduce DashAct, a progressive diagnostic benchmark that investigates
online failures by evaluating the same annotated dashboard interactions
under systematically varied contextual and spatial support.
DashAct contains 348 analytical questions and human-verified interaction
trajectories from 142 real-world dashboards, annotated with action-level
milestones and view- and item-level targets.
These annotations link three diagnostic stages:
\textbf{End-to-End Task Execution} evaluates autonomous interaction and
final-answer correctness while tracking annotated key interactions;
\textbf{Verified-Context Action Prediction} provides the verified preceding
history and current dashboard screenshot to evaluate prediction of the
corresponding next action; and \textbf{Target-Conditioned Grounding}
specifies the intended target to evaluate its visual localization.
The grounding stage further compares full-dashboard localization with
view guidance, masking, and local cropping.
By relating outcomes for the same interactions across these conditions,
DashAct provides diagnostic evidence for interpreting online failures
and identifying potential bottlenecks.

Experiments across general-purpose multimodal models, GUI-specialized agents,
and grounding-specialized models reveal substantial difficulties in
interactive dashboard analysis. Agents often make partial interaction
progress without answering correctly, while action prediction and
fine-grained grounding remain challenging even with additional support.
Models respond differently to contextual and spatial guidance, with
visualization marks posing greater grounding challenges than interface
controls. Qualitative trajectories further illustrate how state tracking,
evidence retention, and interaction feedback affect online behavior.
Guided by these findings, we adapt Qwen3-VL-8B-Instruct through Planner
and Grounder fine-tuning with learned view localization.
Improvements in offline target localization and online Milestone Coverage
demonstrate the practical value of diagnosis-guided agent adaptation.
\section{Related Work}

\paragraph{Visual Data Understanding and Interactive Agents.}
Chart understanding benchmarks evaluate visual perception, reasoning,
and evidence grounding in static visualizations%
~\cite{methani2020plotqa,masry2022chartqa,vogel2025refchartqa,xu2025chartpoint}.
General-purpose GUI agents interact with rendered interfaces%
~\cite{he2024webvoyager,qin2025uitars,wang2025opencua},
while LightVA supports collaborative task planning and execution%
~\cite{zhao2024lightva} and ProactiveVA provides context-aware
analytical assistance~\cite{zhao2025proactiveva}.
VACP makes visualization states and interactions accessible through
an agent-oriented protocol~\cite{stahle2026vacp}.
DashAct examines autonomous screenshot-based interaction with
existing dashboards, where analytical evidence may need to be
revealed and connected through sequences of GUI actions.

\paragraph{Benchmarks for Interactive Agents.}
Existing benchmarks evaluate online task execution%
~\cite{zhou2023webarena,koh2024visualwebarena,xie2024osworld},
next-action prediction~\cite{deng2023mind2web,lu2024weblinx},
intermediate progress~\cite{pan2024webcanvas},
hierarchical planning and action execution~\cite{lin2024videogui},
and visual grounding~\cite{cheng2024seeclick}.
More closely related, DashboardQA evaluates interactive question
answering on real dashboards and analyzes grounding, planning,
and reasoning failures~\cite{kartha2026dashboardqa},
while VizAgentBench assesses coordinated multi-view analysis
through declarative interaction commands~\cite{wu2026vizagentbench}.
DashAct combines online evaluation with reference-action milestone
matching, verified-context action prediction, and target-conditioned
grounding under four visual conditions.
By linking outcomes for the same annotated interactions, DashAct
provides diagnostic evidence for interpreting online failures
and examining models' responses to contextual and spatial support.
\section{Benchmark}

\subsection{Problem Setup}

\paragraph{Interactive Dashboard Tasks.}
We define an interactive dashboard task as a tuple
$\mathcal{T}=(D,q,y)$, where $D$ is an interactive dashboard, $q$ is a
natural-language question, and $y$ is the reference answer. A dashboard
comprises one or more \emph{views}, spatially distinct regions containing
related visual content or controls, such as charts, tables, filter panels,
and toolbars. Within a view, interactive \emph{items} include interface
controls and visualization marks; acting on an item may update its
containing view or other views.
Starting from the initial state of $D$, an agent observes screenshots
and performs GUI actions to acquire evidence before submitting an answer
$\hat{y}$. We represent the trajectory as
$\tau=(o_0,a_0,o_1,\ldots,a_{L-1},o_L,\hat{y})$, where $o_t$ is the
screenshot at step $t$, $a_t$ is the corresponding action, and $L$ is
the number of actions. Actions may update filters, selections, or
coordinated views, requiring evidence to be revealed and connected
across successive interface states.

\paragraph{Task Milestones.}
We represent key interactions in each human-verified reference trajectory
as \emph{action-level milestones}. Each milestone records the reference action, a target description for grounding, and the bounding boxes of the
target item and its containing view.
These annotations provide a common evaluation unit across the three
diagnostic stages, linking online interaction outcomes with next-action
prediction and target grounding. For each task, a directed acyclic graph
(DAG) encodes required precedence relations between milestones, while
independent milestones may be completed in any order.

\subsection{Benchmark Construction}

\paragraph{Dashboard sources and trajectory collection.}
We construct DashAct from real-world multi-view interactive dashboards,
primarily sourced from Tableau Public, covering domains such as retail,
finance, healthcare, and public services. Tableau API metadata is used
only for offline dataset construction and verification.
Three annotators experienced in dashboard analysis explore the selected
dashboards with specific analytical goals and record executable interaction
trajectories, including actions and rendered interface states.
A dashboard may contribute multiple questions and corresponding trajectories.

\paragraph{Annotation drafting and verification.}
Given each recorded trajectory and construction-time metadata, GPT-5 drafts
the question, candidate answer, and preliminary milestone annotations.
Human annotators replay each trajectory to verify reproducibility and
screenshot consistency, validate the question and answer, and review
milestone selection, action types and arguments, target descriptions and
categories, and precedence relations. They annotate or correct view- and
item-level bounding boxes, and a second annotator reviews each completed
annotation.
Confirmed human revisions affected 65.4\% of field instances across
the eight reported annotation fields, with 89.7\% of item- and
view-level bounding boxes revised. 
\paragraph{DashboardQA adaptation.}
We adapt DashboardQA~\cite{kartha2026dashboardqa} to the same
interaction-based setting. We exclude questions requiring information
from multiple dashboards, since we focus on exploratory analysis within
a single dashboard. For each retained question, annotators collect an
executable interaction trajectory and produce milestone and target
annotations using the same human verification procedure as for DashAct.

\subsection{Dataset Statistics}

DashAct contains 348 questions and human-verified interaction trajectories
from 142 real-world dashboards. The trajectories contain an average of 2.15
annotated milestones, ranging from 1 to 11. The adapted DashboardQA set
contains 242 questions and trajectories from 83 dashboards, with an average
of 3.99 annotated milestones.
For initial-state view statistics, we count Tableau sheets with at least
three data rows and exclude trajectories with unavailable summaries.
The mean counts are 9.53 for DashAct and 2.94 for the adapted DashboardQA
set. 

\subsection{Evaluation}

Figure~\ref{fig:teaser} summarizes the three diagnostic stages.
Shared milestone and target annotations link online execution with
action prediction and grounding under different support conditions.
Comparisons across these settings provide diagnostic evidence about
potential bottlenecks.

\paragraph{End-to-End Task Execution.}
Given a task question and a live dashboard in its initial state, the agent
observes screenshots, performs a sequence of interface actions, and submits
a final answer. This stage evaluates the complete interactive analysis
process, including exploring the dashboard, interpreting intermediate
results produced by previous actions, advancing the task across multiple
steps, selecting and grounding actions, and synthesizing the final answer.

We report \emph{Milestone Coverage}, the fraction of annotated milestone
instances matched during online execution, pooled across all task rollouts.
Each milestone is counted at most once within a rollout.
\emph{Answer Accuracy} is the fraction of tasks for which the submitted
answer is semantically consistent with the reference answer.
The two metrics measure progress through annotated interactions and
final-answer correctness, respectively.

We evaluate milestone matching through rule-based comparison of executed
actions with annotated reference interactions. Each action is checked
against the annotated action type and, for coordinate-based reference
actions, whether its target coordinate falls within the human-annotated
target bounding box. The milestone dependency DAG enforces necessary
action ordering: a milestone is counted only after its prerequisite
milestones have been matched, while independent milestones may be
completed in any order. We use a fixed viewport and containerized environment to maintain
consistent rendering. Only 0.03\% of milestone-matching decisions
require manual adjudication, using the same action-type and
target-hit criteria. Final answers are evaluated separately using Gemini 3.1 Flash-Lite for
semantic consistency with reference answers. 

\paragraph{Verified-Context Action Prediction.}
For each annotated milestone whose reference action requires target
coordinates, the model receives the task question, the human-verified
preceding action history, and the full-dashboard screenshot immediately
before the reference action. Each milestone is evaluated independently
from its reference context. The model predicts the next action type
and target coordinates, and the prediction is scored against the
corresponding reference action. This stage evaluates action prediction
and target localization with verified interaction context. Linking
these predictions to the corresponding online milestones enables
instance-level comparisons between the two settings.

All three metrics are computed over evaluated milestones whose reference
actions require target coordinates. We report \emph{Action Type Accuracy},
measuring whether the predicted action type matches the reference type;
\emph{Target Hit Rate}, measuring whether the predicted point falls
within the human-annotated target bounding box; and \emph{Action Success},
requiring both the correct action type and a target hit.
Action arguments other than target coordinates are not evaluated.
\paragraph{Target-Conditioned Grounding.}
For each annotated target, the model receives a human-verified target
description, such as its visible name or role in the interface, together
with a reference dashboard screenshot. The model predicts the coordinates
of the specified target. To examine how view-level guidance and
surrounding visual content affect target localization, we evaluate
four visual conditions:
\emph{Full Image}, which provides the complete dashboard screenshot
as a baseline;
\emph{Red Box}, which marks the target's containing view with a red
bounding box on the full screenshot;
\emph{Gray Mask}, which masks regions outside the containing view
while preserving the full image dimensions; and
\emph{Local Crop}, which provides a cropped image of the containing
view. All conditions use the same target descriptions, annotated
instances, and underlying reference screenshots, enabling paired
comparisons across the four conditions.

We report \emph{Target Hit Rate}, defined as the proportion of predicted
points falling within the human-annotated target bounding box, using
a deterministic point-in-box test. The same evaluation criterion is
applied across all four conditions.
\section{Experiments}

\subsection{Experimental Setup}
\label{sec:experimental_setup}

\paragraph{Evaluated models.}
We evaluate nine core models across all three stages.
The closed-source models are Gemini 3 Flash Preview, GPT-5, and Claude
Sonnet 5; the open-source models are OpAgent-32B~\cite{guo2026opagent},
GUI-Owl-1.5-32B~\cite{xu2026mobileagentv35},
Qwen3-VL-8B-Instruct~\cite{bai2025qwen3vl},
GUI-Owl-1.5-8B~\cite{xu2026mobileagentv35},
OpenCUA-7B~\cite{wang2025opencua}, and
UI-TARS-1.5-7B~\cite{qin2025uitars}.
These include general-purpose multimodal and GUI-specialized models,
with open-source sizes ranging from 7B to 32B.
Stage~3 additionally includes two grounding-specialized models:
UGround-V1-7B~\cite{gou2025uground} and
POINTS-GUI-G~\cite{zhao2026pointsguig}.

\paragraph{Unified evaluation environment.}
All stages use rendered screenshots only, without DOM, HTML,
accessibility-tree, or page-source information.
Stages~1--2 use full-dashboard screenshots, while Stage~3 derives
the four conditions in Section~3.4 from the same reference screenshots.
Screenshots retain their original resolution and aspect ratio.
Models receive semantically equivalent prompts, with coordinate
conventions and output formats following each model's official
recommendations; predicted coordinates are mapped to screenshot
coordinates for execution and evaluation.
For Stage~1, Playwright controls the live dashboard, while the
OpenAI Agents SDK and Hugging Face inference stack provide a unified
interaction interface.
We use a fixed browser viewport and a containerized execution environment.
At each turn, the model receives the current screenshot and its
accumulated history of actions and textual observations.
Each task is run once from its specified Tableau URL and initial state
for at most 20 turns, ending when the model invokes the answer action
or reaches the turn limit.
Environment failures are rerun and are not counted as model failures.

\subsection{Main Results on DashAct}
\label{sec:overall_results}

\begin{table*}[t]
    \centering
    \footnotesize
    \setlength{\tabcolsep}{3.0pt}
    \renewcommand{\arraystretch}{0.98}

    \resizebox{\textwidth}{!}{%
    \begin{tabular}{@{}l*{9}{c}@{}}
        \toprule
        & \multicolumn{2}{c}{Stage 1: End-to-End}
        & \multicolumn{3}{c}{Stage 2: Verified Context}
        & \multicolumn{4}{c}{Stage 3: Grounding} \\
        \cmidrule(lr){2-3}
        \cmidrule(lr){4-6}
        \cmidrule(lr){7-10}
        Model
        & \shortstack{Milestone\\Coverage}
        & \shortstack{Answer\\Accuracy}
        & \shortstack{Action Type\\Accuracy}
        & \shortstack{Target Hit\\Rate}
        & \shortstack{Action\\Success}
        & \shortstack{Full\\Image}
        & \shortstack{Local\\Crop}
        & \shortstack{Gray\\Mask}
        & \shortstack{Red\\Box} \\
        \midrule

        \multicolumn{10}{l}{\textit{Closed-source models}} \\
        Gemini 3 Flash Preview\textsuperscript{$\dagger$}
            & \textbf{70.9} & \textbf{51.7}
            & 69.0 & \textbf{61.3} & \underline{47.9}
            & \textbf{78.4} & \textbf{80.7}
            & \textbf{80.5} & \textbf{78.8} \\
        GPT-5
            & 17.8 & 14.1
            & 72.3 & 8.1 & 7.0
            & 3.7 & 37.7 & 12.2 & 2.2 \\
        Claude Sonnet 5
            & \underline{65.1} & \underline{47.7}
            & \underline{74.1} & \underline{57.8} & \textbf{49.8}
            & \underline{75.2} & 62.0
            & \underline{80.1} & \underline{77.3} \\

        \midrule
        \multicolumn{10}{l}{\textit{Open-source models}} \\
        OpAgent-32B
            & 37.8 & 33.0
            & 54.3 & 37.5 & 34.0
            & 66.7 & 55.5 & 68.7 & 66.1 \\
        GUI-Owl-1.5-32B
            & 28.1 & 17.2
            & 66.7 & 29.3 & 28.7
            & 67.2 & \underline{67.5} & 75.1 & 68.0 \\
        Qwen3-VL-8B-Instruct
            & 16.8 & 11.8
            & 21.7 & 7.4 & 7.2
            & 51.7 & 34.3 & 62.6 & 57.1 \\
        GUI-Owl-1.5-8B
            & 25.0 & 4.6
            & \textbf{89.6} & 23.2 & 22.5
            & 52.1 & 36.3 & 66.9 & 54.8 \\
        OpenCUA-7B
            & 8.7 & 2.6
            & 37.1 & 8.0 & 8.0
            & 37.1 & 49.7 & 36.9 & 21.6 \\
        UI-TARS-1.5-7B
            & 16.2 & 13.5
            & 73.1 & 20.5 & 19.6
            & 57.3 & 40.0 & 64.1 & 61.0 \\

        \bottomrule
    \end{tabular}%
    }

    \vspace{-4pt}
    \caption{Main results on DashAct (\%).
    \textbf{Bold} and \underline{underlining} mark the best and
    second-best results per column, respectively.}
    \label{tab:overall-results}
\end{table*}

Table~\ref{tab:overall-results} presents model performance across
the three evaluation stages on DashAct.

\textbf{End-to-end interactive dashboard analysis remains difficult.}
Gemini 3 Flash Preview achieves the highest Stage~1 Milestone Coverage
and Answer Accuracy, at 70.9\% and 51.7\%, respectively.
Among open-source models, OpAgent-32B leads on both metrics
(37.8\% and 33.0\%), followed by GUI-Owl-1.5-32B
(28.1\% and 17.2\%).
Smaller models achieve at most 13.5\% Answer Accuracy.
Even the strongest model answers only just over half of the tasks correctly.
Analysis by milestone count further shows that average Milestone
Coverage declines only modestly as the number of required milestones
increases, whereas Answer Accuracy drops sharply from one to two
milestones and remains low thereafter.
This divergence highlights the growing gap between partial interaction
progress and successful task resolution in multi-milestone tasks.

\textbf{Providing a verified interaction state is not sufficient for reliable
next-action prediction.}
Claude Sonnet 5 achieves the highest Stage~2 Action Success at 49.8\%,
meaning that even the strongest model correctly predicts the action type
and hits the target on fewer than half of the evaluated actions in a
single attempt. Among the open-source systems, OpAgent-32B performs best
at 34.0\%, but still fails on nearly two-thirds of the milestone actions.
Thus, providing the correct interaction history and dashboard state does
not by itself enable reliable prediction of the required action type
and target.

\textbf{Visual grounding benefits from view guidance but varies
across models.}
Compared with full screenshots, red-box guidance improves Stage~3
grounding accuracy for six of the nine core models, while gray
masking improves eight.
For example, GUI-Owl-1.5-8B improves from 52.1\% to 66.9\%
with gray masking.
Local cropping improves OpenCUA-7B from 37.1\% to 49.7\%,
but reduces Claude Sonnet 5 from 75.2\% to 62.0\%.
Although cropping narrows the visual search area, it also removes
global layout cues and changes image dimensions and target positions.
These changes may interact with models' training distributions
and image preprocessing, potentially offsetting the benefits
of reduced visual search.
These contrasting responses highlight models' sensitivity to
the presentation of spatial support, motivating evaluation
across multiple visual conditions.
We next examine grounding difficulties and diagnostic profiles of
online failures, complemented by qualitative trajectory analysis.

\subsection{Diagnostic Analyses}
\label{sec:diagnostic_analysis}

\begin{table}[t]
    \centering
    \footnotesize
    \setlength{\tabcolsep}{3pt}
    \begin{tabular*}{\columnwidth}
        {@{\extracolsep{\fill}}lccc@{}}
        \toprule
        Model
        & \shortstack{View\\Hit}
        & \shortstack{Target\\Hit}
        & \shortstack{Target Hit\\$\mid$View} \\
        \midrule

        \multicolumn{4}{l}{\textit{Closed-source models}} \\
        Gemini 3 Flash Preview
            & \textbf{84.5} & \textbf{61.3} & \textbf{72.5} \\
        GPT-5
            & 61.8 & 8.1 & 13.1 \\
        Claude Sonnet 5
            & \underline{83.0} & \underline{57.8}
            & \underline{69.6} \\

        \midrule
        \multicolumn{4}{l}{\textit{Open-source models}} \\
        OpAgent-32B
            & 59.1 & 37.5 & 63.5 \\
        GUI-Owl-1.5-32B
            & 58.4 & 29.3 & 50.1 \\
        Qwen3-VL-8B-Instruct
            & 19.7 & 7.4 & 37.7 \\
        GUI-Owl-1.5-8B
            & 65.3 & 23.2 & 35.5 \\
        OpenCUA-7B
            & 20.4 & 8.0 & 39.1 \\
        UI-TARS-1.5-7B
            & 50.3 & 20.5 & 40.8 \\

        \bottomrule
    \end{tabular*}
    \vspace{-4pt}
    \caption{Stage~2 spatial localization (\%).
    Target Hit$\mid$View is conditioned on View Hit.
    Bold and underlining mark the best and second-best results
    per column, respectively.}
    \label{tab:spatial-localization}
\end{table}

\begin{table}[t]
    \centering
    \footnotesize
    \setlength{\tabcolsep}{3pt}
    \begin{tabular*}{\columnwidth}
        {@{\extracolsep{\fill}}lcccc@{}}
        \toprule
        & \multicolumn{2}{c}{Full Image}
        & \multicolumn{2}{c}{Gray Mask} \\
        \cmidrule(lr){2-3}
        \cmidrule(l){4-5}
        Model & Controls & Marks & Controls & Marks \\
        \midrule

        \multicolumn{5}{l}{\textit{Closed-source models}} \\
        Gemini 3 Flash Preview
            & \underline{90.3} & \textbf{64.7}
            & \underline{93.1} & \textbf{66.2} \\
        GPT-5
            & 4.8 & 2.3 & 18.1 & 5.5 \\
        Claude Sonnet 5
            & \textbf{94.4} & \underline{53.4}
            & \textbf{93.9} & \underline{64.4} \\

        \midrule
        \multicolumn{5}{l}{\textit{Open-source models}} \\
        OpAgent-32B
            & 88.5 & 41.7 & 90.8 & 43.4 \\
        GUI-Owl-1.5-32B
            & 88.3 & 43.1 & 92.3 & 55.4 \\
        Qwen3-VL-8B-Instruct
            & 73.2 & 27.1 & 88.3 & 33.2 \\
        GUI-Owl-1.5-8B
            & 76.3 & 24.5 & 89.3 & 41.4 \\
        OpenCUA-7B
            & 52.8 & 19.2 & 47.4 & 24.8 \\
        UI-TARS-1.5-7B
            & 78.8 & 32.7 & 84.9 & 40.2 \\

        \bottomrule
    \end{tabular*}
    \vspace{-4pt}
    \caption{Stage~3 grounding accuracy (\%) for interface controls
    and visualization marks under Full Image and Gray Mask.
    Bold and underlining mark the best and second-best results
    per column, respectively.}
    \label{tab:grounding-target-types}
\end{table}

\subsubsection{Grounding across Spatial Levels and Target Types}
\label{sec:grounding_analysis}

Tables~\ref{tab:spatial-localization} and \ref{tab:grounding-target-types} diagnose grounding failures along two axes: spatial level and target type. Table~\ref{tab:spatial-localization} separates locating the target's containing view from identifying the target within that view, while Table~\ref{tab:grounding-target-types} compares interface controls with visualization marks.

\textbf{View-level and within-view localization reveal distinct
grounding difficulties.}
Gemini 3 Flash Preview achieves the highest View Hit, Target Hit,
and Target Hit$\mid$View, at 84.5\%, 61.3\%, and 72.5\%, respectively.
OpAgent-32B exhibits a different profile from GUI-Owl-1.5-8B:
although it reaches the correct view less often (59.1\% vs.\ 65.3\%),
it achieves a higher Target Hit$\mid$View (63.5\% vs.\ 35.5\%).
By contrast, GPT-5 attains a View Hit of 61.8\% but only
13.1\% Target Hit$\mid$View, showing that reaching the correct
view does not necessarily lead to identifying the intended item.
These profiles distinguish difficulties in locating the relevant
view from those in pinpointing the target within it.

\textbf{Visualization marks are consistently harder to ground
than interface controls.}
Table~\ref{tab:grounding-target-types} compares grounding accuracy
for interface controls and visualization marks under Full Image and
Gray Mask. All nine core models perform better on controls than marks
under both conditions. For example, on full screenshots, Gemini 3
Flash Preview achieves 90.3\% accuracy on controls but only 64.7\%
on marks. An aggregate control--mark gap also remains after
target-size standardization.
This gap is practically important because conventional controls are
often exposed as semantically addressable DOM or accessibility elements,
whereas visualization marks may lack stable, individually addressable
nodes and therefore still require visual grounding. Because these marks
must often be hovered over or selected to reveal values and trigger
coordinated interactions, reliable mark grounding remains a central
requirement for interactive dashboard agents.

\begin{figure}[t]
    \centering
    \includegraphics[width=\columnwidth]{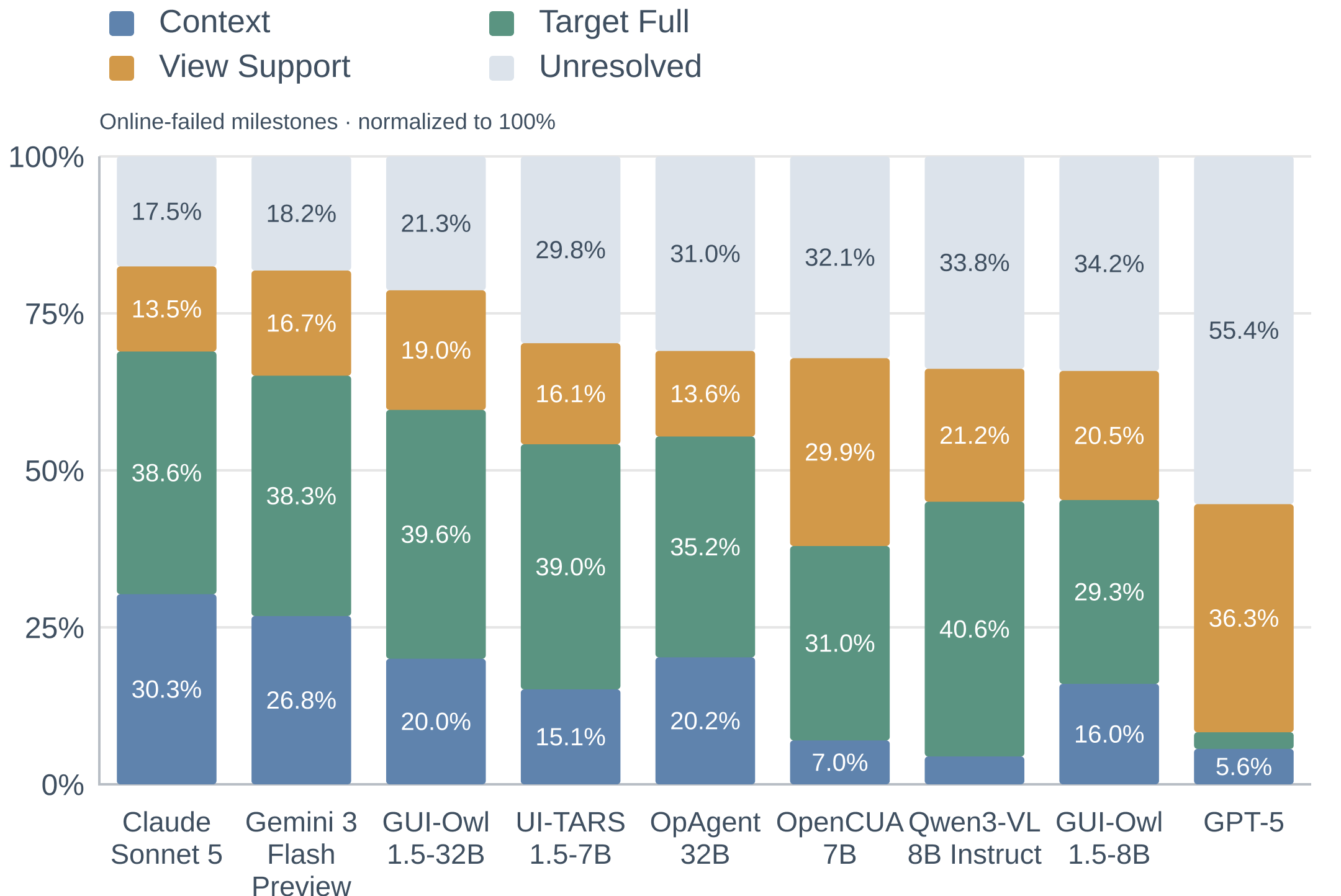}
    \vspace{-2pt}
    \caption{Diagnostic profiles of Stage~1-unmatched milestones.
Percentages are normalized over each model's unmatched milestones
in the shared comparison set.}
    \label{fig:recovery_profiles}
\end{figure}

\subsubsection{Diagnostic Profiles of Online Failures}
\label{sec:recovery_profiles}

Within a shared set of 733 comparable milestones,
Figure~\ref{fig:recovery_profiles} partitions each model's
Stage~1-unmatched milestones into four mutually exclusive categories.
A milestone is assigned to \emph{Context} if it succeeds in Stage~2,
to \emph{Target Full} if it fails in Stage~2 but succeeds under
Stage~3 Full Image, to \emph{View Support} if it fails in both settings
but succeeds under at least one of Gray Mask, Red Box, or Local Crop,
and to \emph{Unresolved} otherwise. Percentages are computed over
each model's Stage~1-unmatched milestones within this shared set.
These priority-ordered categories describe diagnostic outcomes rather
than sequential online recovery. View Support represents an oracle
union across the three view-supported conditions.

\textbf{Some online failures can be correctly predicted with verified context.}
Among Stage~1-unmatched milestones, 30.3\% for Claude Sonnet 5,
26.8\% for Gemini 3 Flash Preview, and 20.0\% for GUI-Owl-1.5-32B
succeed under verified context. These milestones fail during
autonomous execution but succeed when the verified interaction
history and dashboard state are provided, suggesting that some
online failures may involve difficulties in reaching or maintaining
the interaction state required for subsequent actions, rather than
an inability to predict those actions under the correct context.

\textbf{Many remaining targets can be localized when explicitly specified.}
Target Full is the largest additional success category
for eight of the nine models, accounting for 38.3\% of online
failures for Gemini 3 Flash Preview, 40.6\% for
Qwen3-VL-8B-Instruct, and 39.6\% for GUI-Owl-1.5-32B.
These milestones remain unsuccessful under verified-context
action prediction but can be localized when the intended target
is explicitly specified, even without additional view guidance.
This suggests that determining the required action and its target
can present difficulties beyond locating an already specified item.

\textbf{View support enables further localization when full-image grounding fails.}
Among online failures, an additional 36.3\% for GPT-5 and
29.9\% for OpenCUA-7B are correctly localized under at least
one view-supported condition, compared with 16.7\% for
Gemini 3 Flash Preview and 13.5\% for Claude Sonnet 5.
These results suggest that difficulties in locating targets
within full multi-view dashboards can be alleviated by
view-focused visual support, although the benefits vary across
models. Nevertheless, 55.4\% of GPT-5's online failures
remain unresolved across all tested conditions, compared with
21.3\% for GUI-Owl-1.5-32B.

\subsection{Diagnosis-Guided Agent Adaptation}
\label{sec:diagnosis_guided_adaptation}

Guided by the diagnostic findings, we adapt Qwen3-VL-8B-Instruct
into a Planner--Locator--Grounder pipeline.
The fine-tuned Planner predicts actions and target descriptions;
the trained Locator predicts containing views; and the Grounder,
fine-tuned on full screenshots and local crops, predicts target
coordinates. All adapted crops are Locator-predicted, whereas
the base Local Crop condition uses annotated view boxes.
The reference-target comparison therefore evaluates the combined
effect of grounding adaptation and learned view selection.

Table~\ref{tab:diagnosis_guided_adaptation} reports offline results
on 89 instances. With the same Planner and Locator configuration,
Grounder fine-tuning increases Target Hit Rate from 36.0\% to 49.4\%.
Joint GRPO optimization of the Planner and Grounder yields 51.7\%,
while providing reference targets to the SFT pipeline yields 65.2\%.
Online, we compare the SFT pipeline with the base model on the same
35 tasks comprising 86 reference milestones, using matched inference
settings and environment interaction budgets.
Milestone Coverage increases from 14.0\% to 51.2\%, while Answer
Accuracy increases from 5.7\% to 8.6\%.
These results illustrate diagnosis-guided improvements in localization
and online interaction progress, with limited gains in final-answer
accuracy.

\begin{table}[t]
    \centering
    \small
    \setlength{\tabcolsep}{4pt}
    \begin{tabular}{@{}lcc@{}}
        \toprule
        Setting & \multicolumn{2}{c}{Target Hit} \\
        \midrule
        \multicolumn{3}{@{}l}{\textit{Stage 3: reference target descriptions}} \\
        Base, Full Image & \multicolumn{2}{c}{44.9} \\
        Base, Local Crop & \multicolumn{2}{c}{31.5} \\
        Locator + SFT Grounder & \multicolumn{2}{c}{65.2} \\
        \midrule
        \multicolumn{3}{@{}l}{\textit{Stage 2: predicted targets}} \\
        Base, Full Image & \multicolumn{2}{c}{11.2} \\
        Modular, Base Grounder & \multicolumn{2}{c}{36.0} \\
        Modular, SFT Grounder & \multicolumn{2}{c}{49.4} \\
        Joint GRPO (Locator frozen) & \multicolumn{2}{c}{51.7} \\
        \midrule
        \multicolumn{3}{@{}l}{\textit{Stage 1: online execution}} \\
        Agent
        & \shortstack{Milestone\\Coverage}
        & \shortstack{Answer\\Accuracy} \\
        Base, Full Image & 14.0 & 5.7 \\
        SFT modular agent & \textbf{51.2} & \textbf{8.6} \\
        \bottomrule
    \end{tabular}
    \caption{Offline and online results of diagnosis-guided
    Qwen adaptation (\%).}
    \label{tab:diagnosis_guided_adaptation}
\end{table}

\subsubsection{Qualitative Case Analysis}
\label{sec:qualitative_analysis}

To complement the quantitative analyses, we examine representative
trajectories to understand how the identified difficulties manifest
in agent behavior. The cases highlight failures in evidence retention
and prerequisite action selection, as well as the role of interaction
feedback in grounding correction.

\textbf{Agents struggle to track interface state and retain
intermediate evidence across steps.}
In some online rollouts, an action successfully changes the
dashboard, but the agent fails to recognize the resulting
state and repeats the same action, closing a panel or undoing
a completed selection. In other rollouts, agents match all
annotated milestones but still submit incorrect answers,
particularly when tasks require combining values collected
from different dashboard states or performing a final
calculation. These cases highlight two distinct challenges
in multi-step dashboard analysis: tracking interface state
changes and retaining or synthesizing intermediate evidence
for final answer generation.

\textbf{Agents may overlook prerequisite actions needed to access
hidden targets.}
In one OpAgent-32B trajectory, the model continues searching
the visible dashboard for a filter without recognizing that
the filter panel must first be opened. However, when the
panel-opening control is explicitly specified, the model
correctly locates it. When provided with the dashboard state
in which the panel is already open, it also correctly predicts
the subsequent filter interaction.
By comparing these outcomes across diagnostic conditions,
our framework identifies a concrete source of online failure:
the agent overlooks a prerequisite action, preventing it from
reaching the interface state required for subsequent interactions.

\textbf{Interaction feedback can help correct inaccurate grounding.}
In selected Gemini 3 Flash Preview rollouts, an initial hover
lands near the intended visualization mark, but the model
uses the resulting tooltip or visual feedback to adjust its
coordinates until the target is reached.
By contrast, unsuccessful rollouts sometimes repeat inaccurate
actions without using feedback to refine subsequent attempts.
These cases illustrate how interaction feedback can compensate
for imprecise initial grounding, an ability that single-shot
action prediction does not fully capture.
\section{Conclusion}

We introduced DashAct, a progressive diagnostic benchmark with 348
human-verified trajectories from 142 real-world dashboards.
Aligned annotations connect the same interactions across end-to-end
execution, verified-context action prediction, and target-conditioned
grounding, providing diagnostic evidence for interpreting online failures.
Experiments reveal persistent difficulties in action prediction and
fine-grained localization. A Qwen-based adaptation study improves target
localization and online Milestone Coverage, demonstrating the practical
value of diagnosis-guided agent improvement.

\section*{Limitations}

DashAct primarily draws dashboards from Tableau Public, whose
metadata and APIs facilitate dataset construction and verification.
The collected dashboards cover diverse analytical domains, and
construction-time metadata is used only offline; agent evaluation
uses rendered screenshots. Nevertheless, other dashboard platforms
may exhibit different visual conventions and interaction patterns.
Extending the benchmark to additional platforms would broaden its
coverage of dashboard interfaces.

Fine-grained trajectory annotation requires human replay and
verification of milestones, dependencies, and spatial targets.
The current collection therefore covers a bounded range of
analytical workflows. Future extensions could include longer
trajectories with more varied dependency structures, allowing
the diagnostic framework to examine how state tracking and
evidence integration evolve over longer interactions.

Our task formulation focuses on analytical questions within a
single dashboard. Broader workflows may require users to combine
evidence across multiple dashboards. Extending the task scope
to such workflows would provide opportunities to study
cross-dashboard navigation and evidence integration using
the same diagnostic perspective.

\bibliography{references}
\end{document}